\documentclass[preprint,3p]{elsarticle}
\usepackage{amssymb}
\usepackage{amsthm}
\usepackage[mathlines]{lineno}
\usepackage{graphicx}
\usepackage{units}
\usepackage{url}
\usepackage{amsmath}
\usepackage{amsfonts}
\usepackage{mathrsfs}
\usepackage{bm}
\usepackage{textcomp}
\usepackage{subfigure}
\usepackage{multicol}
\usepackage{verbatim}
\usepackage{rotating}
\usepackage[colorlinks,linkcolor=red,citecolor=red]{hyperref}
\usepackage{float}
\usepackage{epsfig}
\usepackage{dcolumn}
\usepackage{bm}
\usepackage{color}
\usepackage{pstricks}
\usepackage{pst-node}
\usepackage{times}
\usepackage{indentfirst}
\usepackage[english]{babel}
\addto{\captionsenglish}{%

}
\journal{Journal}


\newcommand{\XXN}{\Xi^{0}\bar\Xi^{0}}
\newcommand{\XXB}{\Xi^{-}\bar\Xi^{+}}
\newcommand{\XXXB}{\Xi(1530)\bar\Xi(1530)}

\newcommand{\EE}{e^+e^-}

\newcommand{\psp}{\psi(2S)}
\newcommand{\jpsi}{J/\psi}
\newcommand{\ar}{\rightarrow}
\newcommand{\llb}{\Lambda\bar{\Lambda}}

\newcommand{\bfg}{\begin{figure}}
\newcommand{\efg}{\end{figure}}
\newcommand{\bitm}{\begin{itemize}}
\newcommand{\eitm}{\end{itemize}}
\newcommand{\bnum}{\begin{enumerate}}
\newcommand{\enum}{\end{enumerate}}
\newcommand{\btbl}{\begin{table*}}
\newcommand{\etbl}{\end{table*}}
\newcommand{\btbu}{\begin{tabular}}
\newcommand{\etbu}{\end{tabular}}
\newcommand{\bcl}{\begin{center}}
\newcommand{\ecl}{\end{center}}
\newcommand{\bbt}{\bibitem}
\newcommand{\beq}{\begin{equation}}
\newcommand{\eeq}{\end{equation}}
\newcommand{\beqr}{\begin{eqnarray}}
\newcommand{\eeqr}{\end{eqnarray}}

\begin{document}
\begin{frontmatter}
\title{{\bf \boldmath Helicity Amplitude  Analysis of $\jpsi$ and $\psp\ar\XXXB$}}
\author{Xiongfei Wang$^{1}$,   Bo Li$^{2}$, Yuanning Gao$^{3}$, Xinchou Lou$^{1}$
\\
\vspace{0.2cm} {\it
$^{1}$ Institute of High Energy Physics, Beijing 100049, People's Republic of China\\
\vspace{0.2cm} {\it
$^{2}$ Institut de Physique Nucl\'eaire de Lyon, 4, Rue Enrico Fermi, Villeurbanne 69622, France\\
\vspace{0.2cm} {\it
$^{3}$Peking University, Beijing 100871, People's Republic of China
}
}
}
}
\begin{abstract}
We perform a helicity amplitude analysis for the processes of $\EE\ar\jpsi,\psp\ar\XXXB\ar\pi\pi\Xi\bar\Xi\ar 4\pi\llb\ar 6\pi p\bar{p}$. The joint angular distribution for these processes are obtained, which allows a proper estimation of the detection efficiency using helicity amplitude information. In addition, the sensitivities of measurement of hyperon asymmetry decay parameters for these processes are discussed.
\end{abstract}
\end{frontmatter}
\section{Introduction}
Study of charmonium decaying into two-body hyperon anti-hyperon pairs ($Y\bar{Y}$) in $\EE$ annihilation
plays an important role in the validation of the perturbative quantum chromodynamics (pQCD)~\cite{Farrar,Farrar01,Farrar02}.
However, due to the shortage of  helicity amplitude formulae for some $Y\bar{Y}$ final states,  such as  for the processes $\jpsi$, $\psp\ar\XXXB$,  it is hard to give a proper determination of detection efficiency, which affects the precise measurements for their branching fractions, angular distribution, decay parameters, and so on.
In particular, precise measurement of hyperon decay parameters have played an important role in understanding the basic knowledge of particle physics related to the parity violation.  
Although some hyperon decay parameters were measured by the previous experiments with high precision~\cite{PDG2018}, these measurements mainly focused on measuring the $\Lambda$ decay parameters,  $\alpha(\Lambda\ar p\pi^-) = 0.750 \pm 0.009 \pm 0.004$, $\alpha(\bar\Lambda\ar\bar{p}\pi^+) = -0.758 \pm 0.010 \pm 0.007$ and $\alpha(\bar\Lambda\ar\bar{n}\pi^{0}) = -0.692 \pm 0.016 \pm 0.006$ with the uncertainty of $\sim$2.0\%~\cite{Ablikim:2018zay}. For other hyperons,
the decay parameters such as $\alpha(\bar\Xi^+\ar\bar{p}\pi^+)$, $\alpha(\bar\Sigma^+\ar\bar{p}\pi^{0})$, $\alpha(\bar\Sigma^-\ar\bar{p}\pi^{0})$, etc., have not been measured yet, or measured with large uncertainty, such as  $\alpha(\Xi^-\ar\Lambda\pi^-) = -0.458 \pm 0.012$, $\alpha(\Xi^0\ar\Lambda\pi^0) = -0.406 \pm 0.013$ and $\alpha(\Omega^{-}\ar\Lambda K^{-}) = 0.0180 \pm 0.0024$ with the  uncertainties of 3.0\%$\sim$10.0\% level.
To improve the experimental sensitivity or precision, data samples with high statistics are expected.
The experiment for $\EE$ colliders provides us an ideal laboratory to study hyperon property with the improvement of the measured sensitivities, or perform a precise measurement of hyperon and anti-hyperon decay parameters, using charmonium decays  of $\jpsi$ and $\psp$ with low backgrounds and a large data sample.
Thus, it is worth giving an estimation of experimental sensitivity for hyperon decay parameter based on the helicity amplitude formula. 

In this paper, we present a helicity amplitude formula for the process of 
$\EE\ar\jpsi, \psp$$\ar\XXXB$$\ar\pi\pi\Xi\bar\Xi\ar 4\pi\llb\ar 6\pi p\bar{p}$, and obtain the expressions of angular distribution for $\jpsi$ and $\psp\ar\XXXB$ and the $\alpha_{\Lambda}$ and $\alpha_{\Xi}$ decay parameters for $\Lambda\ar p\pi$ and $\Xi\ar\pi\Lambda$ .  
The experimental sensitivities of measurement of the hyperon  asymmetry decay parameter for the corresponding processes mentioned before are discussed based on the largest statistics taken by BESIII experiment~\cite{BES_detector,Ablikim:2017wyh, Ablikim:2016fal}.

\section{Helicity amplitude}
\begin{figure}[!htbp]
\begin{center}
\includegraphics[width=0.65\textwidth]{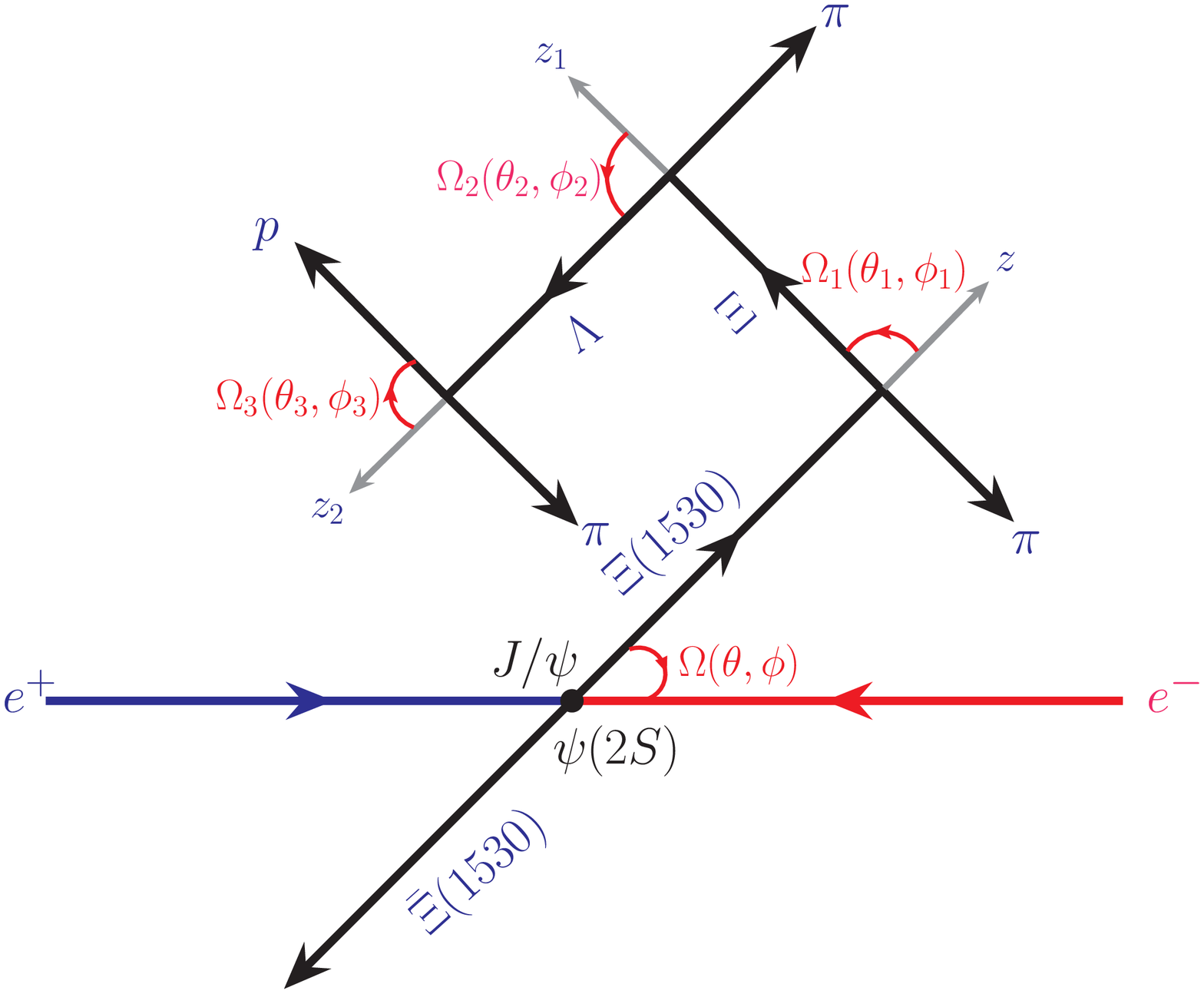}
\caption{Definition of the helicity frame for $\EE\ar\jpsi, \psp$$\ar\XXXB$$\ar\pi\pi\Xi\bar\Xi\ar 4\pi\llb\ar 6\pi p\bar{p}$.}
\label{scatterplot}
\end{center}
\end{figure}

The joint helicity amplitude of the sequential decays for 
\begin{equation}
 \begin{aligned}
\EE&\ar\jpsi, \psi\ar\XXXB\\
&\ar\pi^0\pi^0\Xi^0\bar\Xi^0/\pi^0\pi^0\Xi^-\bar\Xi^+/\pi^+\pi^-\Xi^0\bar\Xi^0/\pi^+\pi^-\Xi^-\bar\Xi^+\\
&\ar\pi^0\pi^0\pi^0\pi^0\llb/\pi^0\pi^0\pi^+\pi^-\llb/\pi^+\pi^+\pi^-\pi^-\llb \\
&\ar\pi^+\pi^-\pi^0\pi^0\pi^0\pi^0p\bar{p}/\pi^+\pi^+\pi^-\pi^-\pi^0\pi^0p\bar{p}/\pi^+\pi^+\pi^+\pi^-\pi^-\pi^-p\bar{p}\\
 \end{aligned}
\end{equation}
can be expressed by
\begin{equation}\label{amplitude}
 \begin{aligned}
|{\cal{M}}|^2 \propto &\sum
\limits_{\substack{
s_{1}, s_{2},\bar{s}_{1},\bar{s}_{2},\\
\xi_{1},\xi_{2},\bar{\xi}_{1},\bar{\xi}_{2},\\
\lambda_{1}, \lambda_{2},\bar\lambda_{1}, \bar\lambda_{2},\lambda_{p},\lambda_{\bar{p}}\\
}}
\rho^{(s_{1} - \bar{s}_{1}, s_{2} - \bar{s}_{2})}
(\theta, \phi)A_{s_{1},\bar{s}_{1}}A^{*}_{s_{2},\bar{s}_{2}}\\
&\times B_{\xi_{1}}B^{*}_{\xi_{2}}\bar{B}_{\bar\xi_{1}}\bar{B}^{*}_{\bar\xi_{2}}{\cal{D}}^{3/2{*}}_{s_{1}, \xi_{1}}(\Omega_{1}){\cal{D}}^{3/2}_{s_{2}, \xi_{2}}(\Omega_{1})
{\cal{D}}^{3/2{*}}_{\bar{s}_{1}, \bar\xi_{1}}(\bar\Omega_{1}){\cal{D}}^{3/2}_{\bar{s}_{2}, \bar\xi_{2}}(\bar\Omega_{1})\\
&\times C_{\lambda_{1}}C^{*}_{\lambda_{2}}\bar{C}_{\bar\lambda_{1}}\bar{C}^{*}_{\bar\lambda_{2}}{\cal{D}}^{1/2{*}}_{\xi_{1}, \lambda_{1}}(\Omega_{2}){\cal{D}}^{1/2}_{\xi_{2}, \lambda_{2}}(\Omega_{2})
{\cal{D}}^{1/2{*}}_{\bar{\xi}_{1}, \bar\lambda_{1}}(\bar\Omega_{2}){\cal{D}}^{1/2}_{\bar{\xi}_{2}, \bar\lambda_{2}}(\bar\Omega_{2})\\
&\times D_{\lambda_{p}}D^{*}_{\lambda_{p}}\bar{D}_{\lambda_{\bar{p}}}\bar{D}^{*}_{\lambda_{\bar{p}}}{\cal{D}}^{1/2{*}}_{\lambda_{1}, \lambda_{p}}(\Omega_{3}){\cal{D}}^{1/2}_{\lambda_{2}, \lambda_{p}}(\Omega_{3})
{\cal{D}}^{1/2{*}}_{\bar{\lambda}_{1}, \lambda_{\bar{p}}}(\bar\Omega_{3}){\cal{D}}^{1/2}_{\bar{\lambda}_{2}, \lambda_{\bar{p}}}(\bar\Omega_{3}),
\end{aligned}
\end{equation}
where, ${\cal{D}}^{J}_{\lambda_{1},\lambda_{2}}(\Omega_{i}) \equiv {\cal{D}}^{J}_{\lambda_{1},\lambda_{2}}(\phi_{i},\theta_{i},0)$, and $\rho^{i,j}(\theta,\phi) = \sum_{k=\pm1}{\cal{D}}^{1}_{k,i}(\Omega){\cal{D}}^{1*}_{k,j}(\Omega)$ is the density matrix~\cite{Richman:1984gh} for the $\jpsi$ and $\psp$ produced in $\EE$ annihilation, $A_{s_{1}, \bar{s}_{2}}$ denotes the helicity amplitude for the processes of $\jpsi$, $\psp\ar\XXXB$, $B_{\xi}$, $C_{\lambda}$ and $D_{\lambda_{p}}$ denote the helicity amplitudes for the processes $\Xi(1530)\ar\pi\Xi$, $\Xi\ar p\pi$ and $\Lambda\ar p\pi$, and $s_{1,2},\bar{s}_{1,2},\xi_{1,2},\bar\xi_{1,2},\lambda_{1,2,p,\bar{p}}, \bar\lambda_{1,2,p,\bar{p}}$ are the helicity values for $\Xi(1530)$, $\Xi$, $\Lambda$ and proton (antiproton), respectively. With the consideration of parity invariance in the processes $\jpsi$, $\psp\ar\XXXB$ and $\Xi(1530)\ar\pi\Xi$, one can get $A_{-s_{1}, -\bar{s}_{2}} = A_{s_{1}, \bar{s}_{2}}$ and $B_{-\xi} = B_{\xi}$ and $\bar{B}_{-\bar\xi} = \bar{B}_{\bar\xi}$.
It should be noted that the definition of helicity frame for $\EE\ar\jpsi, \psp$$\ar\XXXB$$\ar\pi\pi\Xi\bar\Xi\ar 4\pi\llb\ar 6\pi p\bar{p}$ is illustrated in Fig.~\ref{scatterplot}: (1) For $\EE\ar\jpsi, \psp\ar\XXXB$, we choose the $\Xi(1530)$ outgoing direction as the $z$-axis of the $\jpsi$, $\psp$ rest frame, and the direction of the $e^{+}$ beam is at the solid angle $\Omega(\theta, \phi)$; (2) in the subsequent hyperon decays, i.e.  $\Xi(1530)\ar\pi\Xi$, $\Xi\ar\pi\Lambda$ and $\Lambda\ar p\pi$, the solid angle of the daughter particle $\Omega_{i}(\theta_{i},\phi_{i})$ is referred to the individual mother particle  rest frame. The $z_{i}$-axis is taken along the outgoing direction of the individual hyperon in its mother particle rest frame. The helicity frame of anti-hyperon $\bar{Y}_{i}$ is defined with similar to hyperon described by the solid angle $\bar\Omega_{i}(\bar\theta_{i},\bar\phi_{i})$.

By substituting Eq.~(\ref{amplitude22}) into Eq.~(\ref{amplitude}),  one can get 
\begin{equation}\label{amplitude22}
{\cal{D}}^{J}_{m,m^{\prime}}(\theta,\phi) = e^{-im\phi}d^{J}_{m,m^{\prime}}(\theta) ,
\end{equation}
the joint helicity amplitude can be written as 
\begin{equation}\label{amplitude2}
 \begin{aligned}
|{\cal{M}}|^2 \propto &\sum
\limits_{\substack{
s_{1}, s_{2},\bar{s}_{1},\bar{s}_{2},\\
\xi_{1},\xi_{2},\bar{\xi}_{1},\bar{\xi}_{2},\\
\lambda_{1}, \lambda_{2},\bar\lambda_{1}, \bar\lambda_{2},\lambda_{p},\lambda_{\bar{p}}\\
}}
(\sum_{k}d^{1}_{k,s_{1}-\bar{s}_{1}}(\theta)d^{1}_{k, s_{2}-\bar{s}_{2}}(\theta))A_{s_{1},\bar{s}_{1}}A^{*}_{s_{2},\bar{s}_{2}}\\
&\times B_{\xi_{1}}B^{*}_{\xi_{2}}\bar{B}_{\bar\xi_{1}}\bar{B}^{*}_{\bar\xi_{2}}C_{\lambda_{1}}C^{*}_{\lambda_{2}}\bar{C}_{\bar\lambda_{1}}\bar{C}^{*}_{\bar\lambda_{2}}
D_{\lambda_{p}}D^{*}_{\lambda_{p}}\bar{D}_{\lambda_{\bar{p}}}\bar{D}^{*}_{\lambda_{\bar{p}}}\\
&\times e^{i[(s_{1}-s_{2})\phi _{1} + (\bar{s}_{1}-\bar{s}_{2})\bar\phi_{1}]}d^{3/2}_{s_{1}, \xi_{1}}(\theta_{1})d^{3/2}_{s_{2}, \xi_{2}}(\theta_{1})
d^{3/2}_{\bar{s}_{1}, \bar\xi_{1}}(\bar\theta_{1})d^{3/2}_{\bar{s}_{2}, \bar\xi_{2}}(\bar\theta_{1})\\
&\times e^{i[(\xi_{1}-\xi_{2})\phi_{2} + (\bar{\xi}_{1}-\bar{\xi}_{2})\bar\phi_{2}]}d^{1/2}_{\xi_{1}, \lambda_{1}}(\theta_{2})d^{1/2}_{\xi_{2}, \lambda_{2}}(\theta_{2})
d^{1/2}_{\bar{\xi}_{1}, \bar\lambda_{1}}(\bar\theta_{2})d^{1/2}_{\bar{\xi}_{2}, \bar\lambda_{2}}(\bar\theta_{2})\\
&\times e^{i[(\lambda_{1}-\lambda_{2})\phi_{3} + (\bar{\lambda}_{1}-\bar{\lambda}_{2})\bar\phi_{3}]}d^{1/2}_{\lambda_{1}, \lambda_{p}}(\theta_{3})d^{1/2}_{\lambda_{2}, \lambda_{p}}(\theta_{3})
d^{1/2}_{\bar{\lambda}_{1}, \lambda_{\bar{p}}}(\bar\theta_{3})d^{1/2}_{\bar{\lambda}_{2}, \lambda_{\bar{p}}}(\bar\theta_{3}).
\end{aligned}
\end{equation}
Meanwhile, the differential cross section for the corresponding process can be written as
\begin{equation}\label{Omega}
\frac{d\sigma}{d\Omega d\Omega_{1}d\Omega_{2}\Omega_{3}}  \propto  |{\cal{M}}|^{2}.
\end{equation}

By integrating over all azimuthal angles and over the polar angles ($\theta_{1}$, $\bar\theta_{1}$, $\theta_{2}$ and $\bar\theta_{2}$) with the following relations,
\begin{equation}
\int^{2\pi}_{0}e^{i\phi(\lambda-\lambda^{\prime})}d\phi = 2\pi\delta_{\lambda\lambda^{\prime}},
\end{equation}
\begin{equation}
\int^{\pi}_{0}[d^{J}_{m,m^{\prime}}(\theta)]^{2}\sin\theta d\theta  = \frac{2}{2J+1},
\end{equation}
one can reduce amplitude in Eq.~(\ref{amplitude2}) as
\begin{equation}\label{mathematic}
 \begin{aligned}
\frac{d\sigma}{d\cos\theta d\cos\theta_{3}d\cos\bar\theta_{3}}
\propto   \frac{1}{4}&\sum
\limits_{\substack{
s_{1}, \bar{s}_{1},\\
\xi_{1}, \bar{\xi}_{1},\\
\lambda_{1},\bar\lambda_{1},\lambda_{p},\lambda_{\bar{p}}\\
}}
(\sum_{k}[d^{1}_{k, s_{1}-\bar{s}_{1}}(\theta)]^{2})|A_{s_{1},\bar{s}_{1}}|^{2}\\
&\times |B_{\xi_{1}}|^2 |\bar{B}_{\bar\xi_{1}}|^2|C_{\lambda_{1}}|^2 |\bar{C}_{\bar\lambda_{1}}|^2|D_{\lambda_{p}}|^2 |\bar{D}_{\lambda_{\bar{p}}}|^2\\
&\times[d^{1/2}_{\lambda_{1},\lambda_{p}}(\theta_{3})]^2[d^{1/2}_{\bar\lambda_{1}, \lambda_{\bar{p}}}(\bar\theta_{3})]^2.
 \end{aligned}
\end{equation}

Equation~\ref{mathematic} can be further calculated by substituting in the known expressions for the $d$ functions, i.e.
\begin{equation}\label{angular}
 \begin{aligned}
\frac{d\sigma}{d\cos\theta d\cos\theta_{3}d\cos\bar\theta_{3}} 
\propto (1+\alpha_{B}\cos^{2}\theta)(1 + \alpha_{\Lambda}\alpha_{\Xi}\cos\theta_{3})(1 + \alpha_{\bar\Lambda}\alpha_{\bar\Xi}\cos\bar\theta_{3}),
 \end{aligned}
\end{equation}
where
\begin{equation}
\alpha_{B} = \frac{|A_{1/2, -1/2}|^2 + |A_{1/2, 3/2}|^2+ |A_{3/2, 1/2}|^2 - 2(|A_{1/2, 1/2}|^2+|A_{3/2, 3/2}|^2)}{|A_{1/2, -1/2}|^2 + |A_{1/2, 3/2}|^2+ |A_{3/2, 1/2}|^2 + 2(|A_{1/2, 1/2}|^2+|A_{3/2, 3/2}|^2)},
\end{equation}

\begin{equation}
\alpha_{\Xi(1530)} = \frac{|B_{3/2}|^{2} - |B_{-3/2}|^2}{|B_{3/2}|^{2} + |B_{-3/2}|^2} = 0,
\end{equation}
\begin{equation}
\alpha_{\bar\Xi(1530)}  = \frac{|\bar{B}_{3/2}|^{2} - |\bar{B}_{-3/2}|^2}{|\bar{B}_{3/2}|^{2} + |\bar{B}_{-3/2}|^2} = 0,
\end{equation}

\begin{equation}
\alpha_{\Xi} = \frac{|C_{1/2}|^{2} - |C_{-1/2}|^2}{|C_{1/2}|^{2} + |C_{-1/2}|^2},
\end{equation}
\begin{equation}
\alpha_{\bar\Xi} = \frac{|\bar{C}_{1/2}|^{2} - |\bar{C}_{-1/2}|^2}{|\bar{C}_{1/2}|^{2} + |\bar{C}_{-1/2}|^2},
\end{equation}

\begin{equation}
\alpha_{\Lambda} = \frac{|D_{1/2}|^{2} - |D_{-1/2}|^2}{|D_{1/2}|^{2} + |D_{-1/2}|^2},
\end{equation}
\begin{equation}
\alpha_{\bar\Lambda} = \frac{|\bar{D}_{1/2}|^{2} - |\bar{D}_{-1/2}|^2}{|\bar{D}_{1/2}|^{2} + |\bar{D}_{-1/2}|^2},
\end{equation}
by integrating over $\theta$ together with $\bar\theta_{3}$ or $\theta_{3}$, one can get 
\begin{equation}\label{costheta3}
 \begin{aligned} 
\frac{d\sigma}{d\cos\theta_{3}}  \propto (1 + \alpha_{\Lambda}\alpha_{\Xi}\cos\theta_{3}), 
 \end{aligned}
\end{equation}
\begin{equation}\label{costheta3bar}
 \begin{aligned} 
\frac{d\sigma}{d\cos\bar\theta_{3}}  \propto (1 + \alpha_{\bar\Lambda}\alpha_{\bar\Xi}\cos\bar\theta_{3}).
 \end{aligned}
\end{equation}

In the above formulae, the $\alpha_{B}$ is the parameter of angular distribution for the processes $\jpsi$, $\psp\ar\XXXB$, 
$\alpha_{\Xi(1530)/\bar\Xi(1530)}$ is the asymmetry decay parameter for the process of $\Xi(1530)\ar\pi\Xi$, which is calculated to be zero due to the parity conservation,  
$\alpha_{\Xi/\bar\Xi}$ is the asymmetry decay parameter for the process $\Xi\ar\pi\Lambda$,  
and $\alpha_{\Lambda/\bar\Lambda}$ is the asymmetry decay parameter for the process $\Lambda\ar p\pi$.
\label{sec:evt_sel}
\section{Sensitivity}
To estimate the sensitivity of measurement of hyperon decay parameter $\alpha_{\Lambda}\alpha_{\Xi}$ for the processes of $\EE\ar\jpsi, \psp\ar\XXXB\ar\pi\pi\Xi\bar\Xi\ar 4\pi\llb\ar 6\pi p\bar{p}$, we use a maximum likelihood to perform an estimation of experimental sensitivity, where the joint likelihood ${\cal{L}} $ is defined by 
\begin{equation}\label{likelihood}
{\cal{L}} = \prod^{N}_{i=1}f(p_{i},\alpha) = \prod^{N}_{i=1}\frac{|{\cal{M}}(p_{i},\alpha)|^2\epsilon(p_{i})}{\int dp_{i}|{\cal{M}}(p_{i},\alpha)|^2\epsilon(p_{i})},
\end{equation}
and $f(p_{i},\alpha)$ is the probability function of observation of  the $i^{th}$ events with the amplitude  ${\cal{M}}(p_{i},\alpha)$ and momentum of $p_{i}$,  $N$ denotes the number of observed events, $\alpha$ is the measurement parameter and  $\epsilon(p_{i})$ is the detection efficiency. 

The sensitivity of measurement of the parameter $\alpha$ can be defined by:
\begin{equation}\label{alpha}
\delta(\alpha) = \frac{\sqrt{V(\alpha)}}{|\alpha|},
\end{equation}
 where  $V(\alpha)$ is the variance of parameter $\alpha$, and is defined by Eq.(6.19) and Eq.(6.20) in Ref.~\cite{COWAN}, 
 \begin{equation}\label{alpha_lambda}
  \begin{aligned}
V^{-1}(\alpha)  &=E\left[ -\frac{\partial^{2}\ln{\cal{L}}}{\partial\alpha^{2}}\right]\\
&=  N\int -f(p_{i},\alpha) \frac{\partial^{2}\ln f(p_{i},\alpha)}{\partial \alpha^{2}}dp_{i}\\
&= N\int \frac{1}{f(p_{i},\alpha)}\left[\frac{\partial f(p_{i},\alpha)}{\partial \alpha}\right]^{2}dp_{i},
  \end{aligned}
\end{equation}
where $E$ is the expectation value of the estimator $p_{i}$.

From Eq.~(\ref{costheta3}) or Eq.~(\ref{costheta3bar}) into Eq.~(\ref{likelihood}),  one can get
 \begin{equation}\label{function}
   \begin{aligned}
f(\cos\theta_{3},\alpha)&= \frac{1+\alpha\cos\theta_{3}}{\int^{\pi}_{0} (1+\alpha\cos\theta_{3})d\cos\theta_{3}} \\
&= -\frac{1}{2}(1+\alpha\cos\theta_{3}).
  \end{aligned}
\end{equation}

By substituting Eq.~(\ref{function}) into Eq.~(\ref{alpha_lambda}),  one can  further get
  \begin{equation}\label{alpha_function}
  \begin{aligned}
 V^{-1}(\alpha) = N\left(\frac{1}{2\alpha^{3}}\ln\frac{1+\alpha}{1-\alpha} -\frac{1}{\alpha^2}\right).
  \end{aligned}
\end{equation}

Finally, by using Eq.~(\ref{alpha}) and Eq.~(\ref{alpha_function}), the sensitivity of the $\alpha_{\Lambda}\alpha_{\Xi}$ measurement  in the process of $\EE\ar\jpsi, \psp\ar\XXXB\ar\pi\pi\Xi\bar\Xi\ar 4\pi\llb\ar 6\pi p\bar{p}$ can be calculated as
 \begin{equation}
   \begin{aligned}
\delta(\alpha_{\Lambda}\alpha_{\Xi}) = \frac{1}{\sqrt{N}}
\left[ \frac{1}{2\alpha_{\Lambda}\alpha_{\Xi}}\ln\frac{1+\alpha_{\Lambda}\alpha_{\Xi}}{1-\alpha_{\Lambda}\alpha_{\Xi}} -1\right]^{-\frac{1}{2}},
  \end{aligned}
\end{equation}
where the $N$ is calculated by
\begin{equation}
N= N_{\psi}\times\epsilon\times\prod_{i}{\cal{B}}_{i},
\end{equation} 
$N_{\psi}$ is total number of $\jpsi$ and $\psp$ events collected by BESIII experiment~\cite{Ablikim:2017wyh, Ablikim:2016fal}, i.e.  $N^{\rm BESIII}_{\jpsi}$ = $\sim10^{10}$ and $N^{\rm BESIII}_{\psp}$ = $\sim$5.0 $\times 10^{8}$, $\prod_{i}{\cal{B}}_{i}$ stands for the product of branching fraction of the subsequent decays taken from the PDG~\cite{PDG2018}, i.e. 
 \begin{equation}
\prod_{i}{\cal{B}}_{i} ={\cal{B}}[\psi\ar\Xi(1530)\bar\Xi(1530)]\times{\cal{B}}[\Xi(1530)\ar\pi\Xi]\times{\cal{B}}[\Xi\ar\pi\Lambda]\times{\cal{B}}[\Lambda\ar p\pi],
 \end{equation}
estimated roughly to be an order of $\sim$$10^{-4}$ ($\sim$$10^{-5}$) for $\jpsi$ ($\psp$) decay~\cite{wangxf}, 
$\epsilon$ represents the corresponding detection efficiency estimated roughly to be a level of $\sim$10\%  according to Refs.~\cite{Ablikim:2012ena,Ablikim:2016iym,Ablikim:2016sjb}.
Table~\ref{error_ang} summarizes the numerical results for the process  $\EE\ar\jpsi, \psp\ar\XXXB\ar\pi\pi\Xi\bar\Xi\ar 4\pi\llb\ar 6\pi p\bar{p}$.

\begin{table}[!htbp]
\begin{center}
\caption{\label{error_ang}Summary of the expected yields and the expected sensitivity on the $\alpha_{\Lambda}\alpha_{\Xi}$ measurement for the process of $\jpsi$, $\psp\ar\XXXB\ar\pi\pi\Xi\bar\Xi\ar 4\pi\llb\ar 6\pi p\bar{p}$, where
the $\alpha_{\Lambda} = 0.642 \pm 0.013 $,  $\alpha_{\Xi^{-}} = -0.458 \pm 0.012$, $\alpha_{\Xi^{0}} = -0.406 \pm 0.013$ are quoted by the references~\cite{PDG2018, Ablikim:2018zay}, 
i.e. $\alpha_{\Lambda}\alpha_{\Xi^{-}} = 0.347 \pm 0.017$, $\alpha_{\Lambda}\alpha_{\Xi^{0}} = 0.308 \pm 0.018$.}
    \linespread{1.5}
\begin{tabular}{ccccc}  \hline
\multicolumn{1}{l}{Mode} &\multicolumn{1}{c}{Expected yields} &\multicolumn{2}{c}{Sensitivity of $\delta(\alpha_{\Lambda}\alpha_{\Xi})$ (\%)}  \\ \hline
$\jpsi$, $\psp\ar\XXXB$   &$N$    &Measured &$N^{\rm BESIII}_{\jpsi}$($N^{\rm BESIII}_{\psp})$ sample \\ \hline
$\ar\pi^{+}\pi^{-}\XXB$ &100,000 (5,000) &5.0  &$\sim$1.8 ($\sim$8.0)\\
$\ar\pi^{0}\pi^{0}\XXB$ &100,000 (5,000) &5.0  &$\sim$1.8 ($\sim$8.0)\\
$\ar\pi^{+}\pi^{-}\XXN$ &100,000 (5,000) &6.0  &$\sim$2.0 ($\sim$9.2)\\
$\ar\pi^{0}\pi^{0}\XXN$ &100,000 (5,000) &6.0  &$\sim$2.0 ($\sim$9.2)\\ \hline
\end{tabular}
\end{center}
\end{table}

\section{Summary and prospects}
In summary, we perform a helicity amplitude analysis for the process of $\EE\ar\jpsi, \psp\ar\XXXB\ar\pi\pi\Xi\bar\Xi\ar 4\pi\llb\ar 6\pi p\bar{p}$. The formulae of the joint angular distributions for $\jpsi$ and $\psp$ decaying into $\XXXB$ final state are presented, which allows a proper estimation of detection efficiency for the corresponding  processes and determination of the decay parameter of anti-hyperon, as well as extend to the study of other properties of hyperon or anti-hyperon.  
In addition, the sensitivities of measurement of hyperon asymmetry decay parameters for this process are studied based on the formula of helicity amplitude.
Compared with the previous measurements, the asymmetry decay parameter $\alpha_{\Lambda}\alpha_{\Xi}$ can be measured with more precision  at BESIII experiment based on the taken 10 million $\jpsi$ events.

\section{Acknowledgement}
The authors are grateful to the LABEX Lyon Institute of Origins (ANR-10-LABX-0066) of the Universit\'e de Lyon for its financial support within the program ``Investissements d'Avenir'' (ANR-11-IDEX-0007) of the French government operated by the National Research Agency (ANR).
This work is supported in part by 
National Natural Science Foundation of China (NSFC) under Contract No. 11521505;
 Chinese Academy of Science Focused Science Grant; National 1000 Talents Program of China;
China Postdoctoral Science Foundation under Contract No.  2018M630206.


\begin{thebibliography}{**}
\bibitem{wangxf}
According to the BESIII preliminary study, the branching fractions for $J/\psi, \psi(2S)\ar\XXXB$ are measured to be the level of $10^{-4}, 10^{-5}$. And the product of branching fractions for other subsequent decays, i.e. 
${\cal{B}}[\Xi(1530)\ar\pi\Xi]\times{\cal{B}}[\Xi\ar\pi\Lambda]\times{\cal{B}}[\Lambda\ar p\pi]$, is calculated to be a level of $\sim$1.0 according to the PDG.
\end{thebibliography}
\end{document}